\documentclass[12pt]{article}
\usepackage[dvips]{graphics}
\begin{document}

\title{An exact solution for two dimensional wetting\\ with a corrugated wall}
\author{P.\ S.\ Swain and A.\ O.\ Parry \\
Department of Mathematics, Imperial College \\
180 Queen's Gate, London SW7 2BZ \\
 United Kingdom.}
\date{}
\maketitle

\section*{Abstract}
{\small An exact solution of a two dimensional RSOS model of wetting at a corrugated (periodic) wall is found using transfer matrix techniques. In contrast to mean-field analysis of the same problem the wetting transition remains second-order and occurs at a lower temperature than that of the planar system. Comparison with numerical studies and other analytical approaches is made.}

\section*{Introduction}

The study of $d$ dimensional inhomogeneous fluids has traditionally focused on systems which retain some degree of translational invariance \cite{rowl}. Consider, for example, the adsorption of a fluid at a solid substrate modelled as a structureless wall which exerts a one-body potential on the fluid particles. If the interface is planar the equilibrium local number density depends on the normal distance from the wall ($z$, say) and the excess free energy is proportional to the area $A_\pi$ \cite{rowl}. If we suppose that the bulk fluid (infinitely far away from the wall) corresponds to a phase $\alpha$ (e.\ g.\ vapour) attention focuses on the possibility of adsorbing a layer of phase $\beta$ (e.\ g.\ liquid) at the wall-$\alpha$ interface. At bulk two-phase coexistence one of two scenarios is possible. Above the wetting temperature $T_\pi$ the thickness of the adsorbed layer is infinite while below $T_\pi$ it is finite. The wetting phase transition occurs as $T\rightarrow T_\pi^-$ and may be first- or second-order corresponding to the discontinuous or continuous divergence in the height of the $\alpha \beta$ interface (measured from the wall), respectively.

For a non-planar or rough boundary the problem poses new difficulties due to an even greater loss of translational invariance. However, recently a number of independent approaches have been developed to meet such a challenge \cite{us,netz,borgs,stella}. Simple phenomenological arguments \cite{us,netz} indicate that the increase in contact angle $\theta$ \cite{rowl} at the non-planar interface is controlled by the roughness parameter
\begin{equation}
r=\frac{A}{A_\pi} \label{r}
\end{equation}
where $A$ is the surface area of the rough wall. While this has been confirmed at low temperatures $T \ll T_\pi$ using rigorous methods \cite{borgs}, mean-field studies \cite{us} suggest a more complicated scenario in the vicinity of a planar second-order wetting transition temperature. Consider, for example, a corrugated wall described by a local height variable $z_W(x,y) = \sqrt{2} a \sin q x$ where $a$ is the root mean square width of the corrugations and $\frac{2\pi}{q}$ is their period. Analytical analysis based on a Landau-type free energy functional \cite{us} shows that the transition is driven first-order for $a>a^*$, where $a^*$ is {\it smaller} than the bulk correlation length of the $\beta$ phase. This first-order wetting transition occurs at a lower temperature than the second-order phase boundary (denoted $T_\pi$) for the planar system. For $a<a^*$ the phase transition remains second-order and occurs at the {\it same} temperature $T_\pi$, i.\ e.\ there is no wetting temperature shift. 

It is natural to question the robustness of this result when fluctuation effects, occurring outside the mean-field approximation, are included. In $d=3$ (which is the marginal dimension for short-range forces) one would expect the topology of the surface phase diagram to remain largely unchanged \cite{us}. However in $d=2$ fluctuation effects may be extremely strong and modifications are quite possible. Indeed, recent numerical work suggests that the important parameter is not $r$ but $\zeta_S$, the roughness exponent for the substrate. Clearly, for a corrugated wall $\zeta_S=0$  but Giugliarelli and Stella \cite{stella} have considered self-affine substrates  for which the average height fluctuation $a_L$ for a sample of length $L$ scales like $a_L \sim L^{\zeta_S}$. Their numerical results strongly indicate that for $\zeta_S<\zeta$, with $\zeta = \frac{1}{2}$ the thermodynamic roughness exponent \cite{fisher}, the transition remains continuous. However for $\zeta_S>\zeta$ it appears to be first-order. 

In the present article we describe details of an exact transfer matrix analysis of wetting at a corrugated wall which confirms the numerically based suggestion that the transition remains second-order. As we shall see the analysis is considerably more involved than the planar example and requires some care. Importantly we shall show that in contrast to mean-field expectations the wetting transition temperature is reduced by corrugation even if the order of the transition is unaffected.

Consider a wall corrugated so that even sites are $a$ lattice spacings higher than odd sites (see Fig.\ 1). 
\begin{figure}[h]
\begin{center}
\scalebox{0.6}{\includegraphics{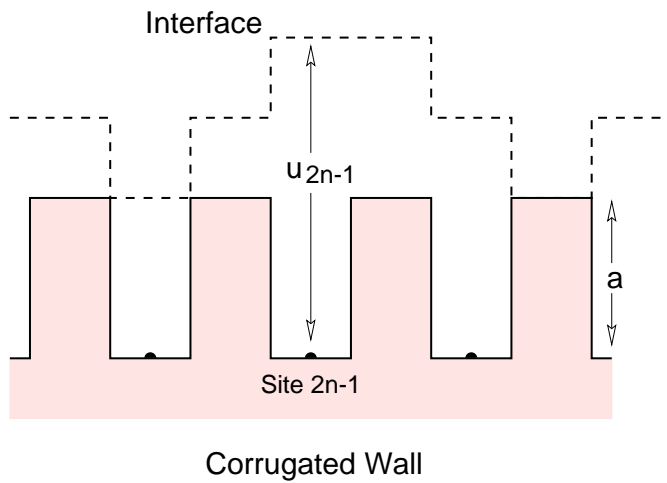}}
\end{center}
\caption{Schematic illustration of a corrugated wall of amplitude $a$. A typical interfacial configuration (dashed line) is also shown.}
\end{figure}
Our approach evokes a restricted solid-on-solid (RSOS) approximation to the semi-infinite Ising model (valid at low temperatures) which has proven to be very profitable for examining wetting transitions \cite{vol14}. Any overhangs of the interface are ignored (as are bubbles of $\alpha$ phase in the $\beta$ phase and vice versa) so that a single valued variable measuring the height of the interface from the wall can be used. To finish our introduction we briefly recall the Hamiltonian for the planar system ($a=0$) and suppose that the wall has $N$ sites and contact potential $W>0$,
\begin{equation}
H[u] = \sum_{n=1}^N \left\{ J|u_{n+1}-u_n| - W \delta_{u_n,0} \right\} \label{Hstandard}
\end{equation}
The variable $u_n$ measure the height of the interface at site $n$ and $J>0$, is the usual coupling parameter which suppresses bending fluctuations. Here periodic boundary conditions imply that $u_{N+1}=u_1$ and with the RSOS constraint adjacent heights are restricted such that $|u_{n+1}-u_n| \in \{0,1 \}$. With this assumption we note that the Boltzmann weight is
\begin{equation}
{\rm e}^{-J|u-u'|/T} = j (\delta_{u',u+1}+\delta_{u',u-1})+\delta_{u,u'} \label{restrict}
\end{equation}
where we have introduced the notation $j={\rm e}^{-J/T}$ (with $0 \le j \le 1$) and set $k_B=1$. The model exhibits a second-order wetting transition when $w\equiv{\rm e}^{W/T}$ satisfies \cite{chalker}
\begin{equation}
w=\frac{1+2j}{1+j} \label{tpi}
\end{equation}
which is the correct low temperature limit of the exact Ising model result \cite{abraham}.

\section*{RSOS model with a corrugated wall}

As adjacent interface configurations are restricted there are only two cases that need to be considered 

\newcounter{d}
\begin{list}{(\roman{d})}{\usecounter{d}}

\item ${\bf a=1.}$ For this value of $a$ the interface can access every site of the wall and the Hamiltonian is 
\begin{equation}
H_1[u] = \sum_{n=1}^N \left\{ J|u_{2n}-u_{2n-1}|+J|u_{2n}-u_{2n+1}| - W( \delta_{u_{2n-1},a-1}+ \delta_{u_{2n},a}) \right\} \label{H1}
\end{equation}

\item ${\bf a>1.}$ For all $a>1$ the interface cannot reach odd sites and still obey $|u_{n+1}-u_n| \in \{0,1 \}$ (see Fig.\ 1). In other words there exists sites that are never accessed.  While this may appear to be an artifact of the RSOS constraint, in an unrestricted interface such configurations would cost enormous bending energy and so would be very rare. Consequently we expect the RSOS approximation to produce little error, particularly for larger values of $a$. The Hamiltonian reads
\begin{equation}
H_{>1}[u] = \sum_{n=1}^N \left\{ J|u_{2n}-u_{2n-1}|+J|u_{2n}-u_{2n+1}| -  W \delta_{u_{2n},a} \right\} \label{H2}
\end{equation}

\end{list}

It is in fact possible to account for both these scenarios with only one Hamiltonian
\begin{equation}
H[u] = \sum_{n=1}^N \left\{ J|u_{2n}-u_{2n-1}|+J|u_{2n}-u_{2n+1}| - W^* \delta_{u_{2n-1},a-1} - W \delta_{u_{2n},a} \right\} \label{H}
\end{equation}
where here there are $2N$ sites along the wall, $u_{2n} \ge a$, $u_{2n-1} \ge a-1$ and periodic boundary conditions are still imposed ($u_{2N+1}=u_1$).  For $a=1$ and $a>1$ the effective variable $W^*$ takes the values $W$ and $0$, respectively.

The partition function ${\cal Z}$ may be written as
\begin{equation}
{\cal Z} = \sum_{\{ u\}} \prod_{n=1}^N {\rm e}^{-f(u_{2n-1},u_{2n},u_{2n+1})} \label{z}
\end{equation}
where $f$ is defined to be
\begin{equation}
f(u,u',v) = J|u-u'|+J|u'-v|-\frac{1}{2}W^* \left( \delta_{u,a-1}+\delta_{v,a-1} \right)-W \delta_{u',a} 
\end{equation}
To simplify (\ref{z}) we follow\cite{scalapino} and write
\begin{equation}
{\cal Z} = \sum_{\{ u\},p} \delta_{p,u_1} {\rm e}^{-f(u_{2N-1},u_{2N},u_p)} \cdots {\rm e}^{-f(u_1,u_2,u_3)}
\end{equation}
Introducing a complete set of normalised eigenfunctions $\psi^{(i)}_n$ such that
\begin{equation}
\sum_i \bar{\psi}^{(i)}_n \psi^{(i)}_m = \delta_{n,m} \label{complete}
\end{equation}
with the over-bar denoting complex conjugation, allows ${\cal Z}$ to be rewritten
\begin{equation}
{\cal Z} = \sum_{i,p,\{ u\} } \bar{\psi}^{(i)}_p {\rm e}^{-f(u_{2N-1},u_{2N},u_p)} \cdots {\rm e}^{-f(u_1,u_2,u_3)} \psi^{(i)}_{u_1} \label{inter}
\end{equation}
If the eigenfunctions are chosen to satisfy
\begin{equation}
\sum_{\stackrel{\scriptstyle u=a-1}{u'=a}}^{\infty} {\rm e}^{-f(u,u',v)} \psi^{(i)}_u = \lambda^{(i)} \psi^{(i)}_v \label{eigen}
\end{equation}
then by using (\ref{eigen}) repeatedly, (\ref{inter}) becomes
\begin{eqnarray}
{\cal Z} & = & \sum_{i,p} \bar{\psi}^{(i)}_p \psi^{(i)}_p \left( \lambda^{(i)} \right)^N \nonumber \\
& = & \sum_i \left( \lambda^{(i)} \right)^N \nonumber \\
& = &  \left( \lambda^{(0)} \right)^N \left[ 1 + \left( \frac{\lambda^{(1)}}{\lambda^{(0)}} \right)^N + \cdots \right] \label{lambdas}
\end{eqnarray}
where $\lambda^{(0)}$ is the maximum eigenvalue. In the thermodynamic limit $N \rightarrow \infty$ the free energy density ${\sf f}$ reduces to
\begin{eqnarray}
{\sf f} & = & - \stackrel{\rm lim}{\scriptscriptstyle N \rightarrow \infty} \frac{ \log Z}{2N} \nonumber \\
& = & - \log \left( \sqrt{\lambda^{(0)}} \right) \label{F}
\end{eqnarray}
where we have used (\ref{lambdas}). As usual the calculation of the free energy reduces to the problem of finding the largest eigenvalue of the matrix equation (\ref{eigen}).

\section*{Solution of the eigenvalue equation}

Consider (\ref{eigen}) when $i=0$
\begin{equation}
\sum_{\stackrel{\scriptstyle u=a-1}{u'=a}} {\rm e}^{-f(u,u',v)} \psi_u = \lambda \psi_v \label{e}
\end{equation}
where the superscripts have been dropped for convenience. Using the Boltzmann factor (\ref{restrict}) this reduces to 
\begin{eqnarray}
& {\displaystyle \sum^\infty_{\stackrel{\scriptstyle u=a-1}{u'=a}}} \left[ j(\delta_{u,u'+1}+\delta_{u,u'-1})+\delta_{u,u'} \right] \left[ j(\delta_{u',v+1}+\delta_{u',v-1})+\delta_{u',v} \right] & \nonumber \\
& \times \sqrt{w^*}^{(\delta_{u,a-1}+\delta_{v,a-1})} w^{\delta_{u',a}} \psi_u = \lambda \psi_v&  \label{res}
\end{eqnarray}
For values of $v>a+1$, $\psi_v$ is beyond the influence of the wall and (\ref{res}) becomes
\begin{equation}
j^2 \psi_{v-2} + 2j \psi_{v-1} +(1+2j^2-\lambda)\psi_{v} + 2j\psi_{v+1} + j^2 \psi_{v+2}=0  \label{free}
\end{equation}
The linearity suggests the ansatz $\psi_{v} = t^v$ for some constant $t$. This gives
\begin{equation}
t^4+2j^{-1}t^3+(j^{-2}+2-\lambda j^{-2})t^2 + 2 j^{-1}t+1 =0
\end{equation}
which in turn simplifies to
\begin{equation}
\left[ t^2+j^{-1} ( 1-\sqrt{\lambda})t+1 \right] \left[t^2+j^{-1}(1+\sqrt{\lambda})t+1 \right] =0 \label{factor}
\end{equation}
Denoting the roots $t_1,t_2$ and $t_3,t_4$ for each bracket we see that
\begin{equation} 
\begin{array}{lcl}
t_1 t_2 =1 & \hspace*{5mm} {\rm and} \hspace*{5mm} & t_1 + t_2 = -j^{-1}(1-\sqrt{\lambda})  \\
t_3 t_4 =1 & {\rm and} & t_3 + t_4 = -j^{-1}(1+\sqrt{\lambda}) \label{fs}
\end{array}
\end{equation}
with (\ref{free}) having a general solution
\begin{equation}
\psi_v = A_0 t_1^v + B_0 t_2^v + A_1 t_3^v + B_1 t_4^v \label{sofar}
\end{equation}
for some constants $A_0,A_1,B_0$ and $B_1$. However we expect the eigenfunction $\psi_v$ to be finite as $v \rightarrow \infty$ and so if the $t_i$ are real we can choose without loss of generality, see (\ref{fs}), $|t_1|,|t_3| \le 1$ and $|t_2|,|t_4| \ge 1$, which implies that $B_0$ and $B_1$ must vanish in (\ref{sofar}). Notice that this is not true if the roots are complex and then $t_1 = \bar{t_2} = {\rm e}^{i\theta}$ for some $\theta$. By considering the discriminants of (\ref{factor}) it is straight forward to show 
\begin{equation}
\begin{array}{lclcl}
t_1 = - {\rm e}^{-\bar{\mu}} & ; &  \sqrt{\lambda} = 1-2j \cosh \bar{\mu} & ; &  \sqrt{\lambda} \le 1-2j \nonumber \\
t_1 = {\rm e}^{i \theta} & ; & \sqrt{\lambda} = 1 +2j \cos \theta & ; & 1-2j<\sqrt{\lambda}<1+2j \nonumber \\
t_1 = {\rm e}^{-\mu} &  ; & \sqrt{\lambda} = 1+2j \cosh \mu & ; & \sqrt{\lambda} \ge1+2j \label{t1}
\end{array}
\end{equation}
for some $\mu,\bar{\mu}$ and $\theta$, and 
\begin{equation}
\begin{array}{lclcl}
t_3 = - {\rm e}^{-\mu_1} & ; &  \sqrt{\lambda} = 2j \cosh \mu_1 -1 & ; &  \sqrt{\lambda} \ge 2j-1 \nonumber \\
t_1 = {\rm e}^{i \theta_1} & ; & \sqrt{\lambda} = 2j \cos \theta_1-1 & ; & \sqrt{\lambda}<2j-1  \label{t3}
\end{array}
\end{equation}
for some $\mu_1$ and $\theta_1$. The eigenfunction $\psi_v$ is
\begin{equation}
\psi_v = A_0 t_1^v + A_1 t_3^v \label{soln}
\end{equation}
for $v>a+1$. In the thermodynamic limit only the maximum eigenvalue contributes to the free energy and so for low temperatures (for which the eigenfunction must correspond to a bound state) we have $\sqrt{\lambda}>1+2j>2j-1$ and (\ref{soln}) is valid for
\begin{eqnarray}
t_1 & = & {\rm e}^{-\mu} = \frac{1}{2} \left\{  j^{-1} ( \sqrt{\lambda}-1)-\sqrt{j^{-2}(\sqrt{\lambda}-1)^2-4} \right\} \label{t1real} \\
t_3 & = & -{\rm e}^{-\mu_1} = -\frac{1}{2} \left\{ j^{-1}(\sqrt{\lambda}+1)-\sqrt{j^{-2}(\sqrt{\lambda}+1)^2-4} \right\}  \label{t3real}
\end{eqnarray}
Wetting will occur when any $t_i$ turns complex. From (\ref{t1}) and (\ref{t3}) this will happen first for $t_1$ when $\sqrt{\lambda}$ decreases to
\begin{equation}
\sqrt{\lambda} = 1+ 2j  \label{wet}
\end{equation}
To deal with the boundary conditions, i.\ e.\ (\ref{res}) for values of $v<a+2$, we extend the solution (\ref{soln}) to the range $v\ge a+1$ and impose
\begin{equation}
\psi_v = 0 \mbox{ \ \ \ \ \ for all $v<a-1$}
\end{equation}
due to the RSOS constraint. Then (\ref{res}) for $v=a-1,a,a+1$ and $a+2$ gives a set of equations for $\psi_{a-1}, \psi_a, A_0$ and $A_1$. These are best written in matrix form
\small
\begin{eqnarray}
& \left( \begin{array}{llll}
j^2 w w^*-\lambda & jw \sqrt{w^*} & j^2 t_1^2 w \sqrt{w^*} & j^2 t_3^2 w \sqrt{w^*} \\
jw \sqrt{w^*} & j^2 +w-\lambda & jt^2_1(1+jt_1+w) & j t_3^2(1+jt_3+w) \\
j^2w \sqrt{w^*} & j(1+w) & jt_1[(w-1)jt_1-2]-j^2 & jt_3[(w-1)jt_3-2]-j^2 \\
0 & j^2 & -j^2t_1 & -j^2 t_3 
\end{array} \right) & \nonumber \\
& \times \left( \begin{array}{c}
\psi_{a-1} \\ \psi_a \\ A_0 \\ A_1
\end{array} \right)
= 0 & \label{lam0}
\end{eqnarray}
\normalsize
To have non-trivial solutions the matrix, ${\cal M}$ say, must have zero determinant
\begin{equation}
\det {\cal M} = 0 \label{lam}
\end{equation}
which is a (complicated) equation for $\lambda$ in terms of $t_1$ and $t_3$.  From here one can of course proceed analytically but the task is laborious and ideally suited for computer algebra packages. Using MATHEMATICA $\lambda$ can be obtained exactly and then used to solve (\ref{lam0}), remembering the normalization condition $\sum_v |\psi_v|^2=1$. The resulting expressions for $\psi_{a-1}, \psi_a, A_0$ and $A_1$ are very long and not particularly illuminating.

However, exactly at wetting (\ref{wet}) holds and it is easy to demonstrate from (\ref{t1real}) and (\ref{t3real}) that
\begin{eqnarray}
t_1 & = & 1 \\
t_3 & = & -(j^{-1}+1)+j^{-1} \sqrt{1+2j} 
\end{eqnarray}
and defining 
\begin{equation}
q=\sqrt{1+2j} \mbox{\ \ \ \ \ $\Rightarrow 1<q<\sqrt{3}$} \label{q}
\end{equation}
equation (\ref{lam}) can be shown to become
\begin{equation}
(q^2-1)^2 w^*w +2q(1+q-q^2+q^3)w=4q^4 \label{TW2}
\end{equation}
which gives the wetting temperature. 

As an aside, we point out that our method of solution is quite versatile and can be very simply adapted to study the situation considered by Nechaev and Zhang \cite{nz}. These authors have a periodic potential at a planar wall --- $W$ taking the values $W_0$ and $W_1$ at odd and even sites respectively. Their Hamiltonian is
\begin{equation}
H[u] = \sum_{n=1}^N \left\{ J|u_{2n}-u_{2n-1}|+J|u_{2n}-u_{2n+1}| - W_0 \delta_{u_{2n-1},0} - W_1 \delta_{u_{2n},0} \right\}
\end{equation}
which is very similar to (\ref{H}). Indeed, away from the wall (\ref{free}) still holds and only the elements in the matrix ${\cal M}$ change. Wetting again occurs when $\sqrt{\lambda}=1+2j$ and using the same techniques as before the wetting temperature is found to obey
\begin{equation}
4q^4+(1+q^2)(q^2-2q-1)w_0 w_1+2q^3(1-q)(w_0+w_1)=0
\end{equation}
where $w_0={\rm e}^{W_0/T}$ and $w_1 = {\rm e}^{W_1/T}$. With a little effort this can be shown to be equivalent to equation (14) in \cite{nz} which provides a useful check for the validity of our method.

\section*{Comparison of wetting temperatures}

We consider the two possible wall configurations separately 

\newcounter{c}
\begin{list}{(\roman{c})}{\usecounter{c}}

\item ${\bf a=1.}$ In this case $w^*=w$ and (\ref{TW2}) is a quadratic equation in $w$. Hence we arrive at
\begin{equation}
w_1(q_1) = \frac{-q_1-q_1^2+q_1^3-q_1^4+q_1 \sqrt{1+2q_1+3q_1^2-5q_1^4-2q_1^5+5q_1^6}}{(q_1^2-1)^2} \label{w1}
\end{equation}
for the wetting phase boundary $w_1={\rm e}^{W/T_1}$. For fixed values of $W$ and $J$ the wetting temperature $T_1$ is reduced from its value $T_\pi$ in the planar case, which from (\ref{tpi}) satisfies
\begin{equation}
w_\pi = \frac{2q_\pi^2}{1+q_\pi^2} \label{wpi}
\end{equation}
To see this first assume 
\begin{equation}
T_1>T_\pi \label{assump}
\end{equation}
Then since $w={\rm e}^{W/T}$ and $j={\rm e}^{-J/T}$ we have 
\begin{eqnarray} 
w_\pi& >& w_1 \label{ineq1} \\
q_1 & >& q_\pi  \label{ineq2}
\end{eqnarray} 
as $j_1>j_\pi$. Using (\ref{w1}) and (\ref{wpi}), equation (\ref{ineq1}) implies the inequality
\begin{equation}
(1-q_1)q_1(1+q_1)^2 q^2_\pi + (1+q_1-q_1^2-q_1^3+q_1^4)q^4_\pi  > q_1^4 
\end{equation}
or (as $q_1^4>q^4_\pi$)
\begin{equation}
(1-q_1)q_1(1+q_1)^2+q_1(1-q_1)^2(1+q_1)q^2_\pi > 0 
\end{equation}
This then gives
\begin{equation}
1+q_1+(1-q_1)q^2_\pi< 0 
\end{equation}
but $q^2_\pi<3$ from (\ref{q}) and so
\begin{equation}
1+q_1+3(1-q_1) <0 
\end{equation}
indicating that $q_1>2$. However, this is a contradiction --- from (\ref{q}) we have $q<\sqrt{3}$ and hence the assumption (\ref{assump}) is false. Thus $T_1$ is lower than $T_\pi$.

\item  ${\bf a>1.}$ Now $W^*=0$ implying $w^*=1$ and (\ref{TW2}) gives
\begin{equation}
w_2(q_2)=\frac{4 q_2^4}{1+2q_2-2q_2^3+3q_2^4} \label{w2}
\end{equation}
where we write $w_2 = {\rm e}^{W /T_2}$ defining the wetting temperature $T_2$ for $a>1$. The shift of the wetting phase boundary can be seen following a similar method to that given above.

First assume 
\begin{equation}
T_2>T_1 \label{assump2}
\end{equation}
which then implies, 
\begin{eqnarray}
w_1(q_1)-w_2(q_2) & > & 0 \label{w12} \\
q_2 &>& q_1 >1 \label{q12}
\end{eqnarray}
and recall from (\ref{q}), $1<q_2<\sqrt{3}$. It is straight forward to see, via (\ref{w2}), that $w'_2(q)>0$ for all $1<q<\sqrt{3}$, and $w_2(1)=1$. Hence (\ref{q12}) gives  $0<w_2(q_1)<w_2(q_2)$. Then, defining $w(q) \equiv w_1(q)-w_2(q)$, (\ref{w12}) becomes
\begin{equation}
w(q_1) > w_1(q_1)-w_2(q_2)>0 \mbox{\ \ \ \ \ for $1<q_1<\sqrt{3}$} \label{wcompare}
\end{equation}
However one can also show from (\ref{w1}) and (\ref{w2}) that $w'(q)<0$ for $1<q<\sqrt{3}$, and $w(1)=0$. Hence $w(q)<0$ for this range of $q$ which contradicts (\ref{wcompare}) based on the assumption (\ref{assump2}).
\end{list}

In summary we have proven that
\begin{equation}
T_2<T_1<T_\pi \label{temps}
\end{equation}
so that the larger the magnitude of corrugations of the wall the further reduced is the wetting temperature. Inequality (\ref{temps}) is simply explained using entropy arguments in the conclusion.

Finally we turn to a calculation of the critical exponents characterising the transition in the non-planar system. First consider the mean interface position above the wall defined by
\begin{equation}
\ell \equiv \langle u \rangle = \frac{\sum_{ \{ u \} } u {\rm e}^{-H[u]}}{\sum_{ \{ u \} } {\rm e}^{-H[u]}}
\end{equation}
In the thermodynamic limit this can be written (for any odd numbered site)
\begin{equation}
\langle u \rangle_{\rm odd}=\frac{\sum_{u=a-1}^\infty u |\psi_u|^2}{\sum_u |\psi_u|^2} \label{uaverage}
\end{equation}
where we have expressed the wave-function normalization explicitly. The sums in (\ref{uaverage}) are simple geometric progressions and are easily evaluated.  Expanding the result around the critical point $\mu=0$ shows the leading-order contribution to be
\begin{equation}
\ell \approx \frac{1}{2\mu} \label{beta}
\end{equation}
where  $\mu \sim T_i-T$ and so the adsorption critical exponent $\beta_s=1$ identical to the planar case \cite{vol14}. Clearly the phase transition remains continuous and falls in the same universality class as the planar system.

Similar remarks apply to the singular contribution to the excess free energy and transverse correlation length $\xi_\parallel$. For example, for the free energy density we find
\begin{eqnarray}
{\sf f} &=& -\log \sqrt{\lambda}  \nonumber \\
&\simeq&  -\log(1+2j) - \frac{j \mu^2}{1+2j}
\end{eqnarray}
for small $\mu$, identifying the specific heat exponent $\alpha_s=0$. Again, the exponent describing the divergence of the correlation length $\nu_\parallel=2$ as expected \cite{vol14}.

\section*{Conclusion}
We have presented an exact analytic solution to the problem of wetting in two dimensions with a corrugated wall. The corrugations can be of arbitrary height without the transition losing its second-order character. The critical exponents governing thermodynamic quantities as the phase boundary is approached are unchanged from the planar case although the wetting temperature itself is reduced. These results contrast sharply with mean-field theory \cite{us} but support numerical studies in $d=2$ with self-affine (rough) walls \cite{stella}. In fact both the $d=2$ and mean-field result suggest that by including thermal fluctuations second-order wetting transitions become roughness-induced first-order if the width of the wall (as measured by a roughness exponent $\zeta_S$ for $d<3$ and length scale $a$ for $d>3$) is larger than the analogous quantity for the fluid ($\alpha \beta$) interface.

Our explicit calculations illustrate that the transition temperature is reduced by wall corrugation and to complete our article we make some comment on the generality of the results. Below the upper critical dimension wetting transitions are bought about by entropy winning its competition with internal energy in the free energy of the interface. For a corrugated wall the internal energy of a bound state is less negative than the planar case because either there are fewer points of contact with the substrate if the $\alpha \beta$ interface is flat or an increased bending energy if it leaves the plane to reach the contact points. Consequently entropy, which is hardly affected by the corrugation, needs to overcome an internal energy of reduced magnitude implying that the wetting temperature is decreased. 

With this general argument we expect that the RSOS result described here captures all the essential physics of the full Ising model calculation (at least at low temperatures). While an Ising model solution would be very welcome the transfer matrix analysis would be much more complex and may be prohibitively difficult. 

We acknowledge financial support from the Engineering and Physical Science Research Council, United Kingdom.

\end{document}